\documentclass[conference]{IEEEtran}
\IEEEoverridecommandlockouts
\usepackage{cite}
\usepackage{amsmath,amssymb,amsfonts}

\usepackage{algorithm}
\usepackage{algpseudocode}

\usepackage{booktabs}
\usepackage{graphicx}
\usepackage{textcomp}
\usepackage{xcolor}
\def\BibTeX{{\rm B\kern-.05em{\sc i\kern-.025em b}\kern-.08em
    T\kern-.1667em\lower.7ex\hbox{E}\kern-.125emX}}
\begin{document}

\title{Primitive-Driven Acceleration of Hyperdimensional Computing for Real-Time Image Classification\\
}

\author{
\IEEEauthorblockN{Dhruv Parikh, Jebacyril Arockiaraj, and Viktor Prasanna}
\IEEEauthorblockA{
\textit{University of Southern California}, Los Angeles, CA, USA \\
\{dhruvash, arockiar, prasanna\}@usc.edu}
}

\maketitle

\begin{abstract}
Hyperdimensional Computing (HDC) represents data using extremely high-dimensional, low-precision vectors---termed hypervectors (HVs)---and performs learning and inference through lightweight, noise-tolerant operations. However, the high dimensionality, sparsity, and repeated data movement involved in HDC make these computations difficult to accelerate efficiently on conventional processors. As a result, executing core HDC operations---binding, permutation, bundling, and similarity search---on CPUs or GPUs often leads to suboptimal utilization, memory bottlenecks, and limits on real-time performance.

In this paper, our contributions are two-fold. First, we develop an image-encoding algorithm that, similar in spirit to convolutional neural networks, maps local image patches to hypervectors enriched with spatial information. These patch-level hypervectors are then merged into a global representation using the fundamental HDC operations, enabling spatially sensitive and robust image encoding. This encoder achieves \textbf{95.67\%} accuracy on MNIST and \textbf{85.14\%} on Fashion-MNIST, outperforming prior HDC-based image encoders. Second, we design an end-to-end accelerator that implements these compute operations on an FPGA through a pipelined architecture that exploits parallelism both across the hypervector dimensionality and across the set of image patches. Our Alveo U280 implementation delivers \textbf{0.09\,ms} inference latency, achieving up to \textbf{1300$\times$} and \textbf{60$\times$} speedup over state-of-the-art CPU and GPU baselines, respectively.
\end{abstract}


\section{Introduction}
\label{sec:intro}

Hyperdimensional Computing (HDC) represents data using high-dimensional, low-precision hypervectors and manipulates them with a small set of algebraic primitives such as bundling, binding, and permutation~\cite{kanerva,survey-two-parter-part-I,survey-two-parter-part-II,survey-hdc-stoch-framework}. These operations enable robust, noise-tolerant learning and lend themselves naturally to bit-level and word-level parallelism, making HDC attractive for edge and resource-constrained platforms~\cite{survey-hdc-edge-intel-progress}. HDC has been successfully applied to biosignals, speech, DNA analytics, and human activity recognition~\cite{trad-hdc-1,trad-hdc-2,trad-hdc-3-imani-voice-hd,trad-hdc-4-imani-dna,trad-hdc-5-imani-hierarchical-hd,static-encoding-1-programmable-hdc}, often achieving competitive accuracy with significantly lower computational cost.

Recent work has focused on improving hypervector design and encoding strategies to bridge the accuracy gap with deep learning while preserving HDC’s efficiency. This includes hardware-aware hypervector construction~\cite{adv-encoding-1-hv-design-eff-hdc}, encoding for binarized images~\cite{adv-encoding-2-encoding-binarized-img}, static optimization of HDC computations~\cite{adv-encoding-3-hardware-aware-static-opt}, and ultra-efficient engines tailored to IoT workloads~\cite{adv-encoding-4-tiny-hd}. In parallel, trainable and adaptive encoders~\cite{trainableHD,trainable-hd-old-version,multi-mani-hd-imani} and algorithm–hardware co-design techniques~\cite{hetero-algorithm-hardware-co-design} have demonstrated that modest learning capacity and task-aware encodings can substantially improve classification accuracy without abandoning HDC’s lightweight primitives.

However, image-specific HDC methods remain comparatively underexplored. Existing work on image classification with HDC often treats images as unordered collections of pixels or handcrafted descriptors, or emphasizes privacy and energy efficiency rather than spatially structured representations~\cite{static-encoding-8-distri-HD,static-encoding-9-hdc-framework-image-descriptors-cvpr,rosing-vision-hd,adv-encoding-2-encoding-binarized-img}. These approaches typically lack an explicit notion of patch-level receptive fields and spatial permutation analogous to convolutional neural networks, and they seldom study how such encoders map to concrete hardware implementations for real-time deployment. As a result, there is a gap between the algorithmic potential of HDC for image data and its realization as a low-latency, hardware-efficient pipeline.

On the hardware side, a growing body of work accelerates generic HDC workloads on CPUs, GPUs, FPGAs, and processing-in-memory (PIM) substrates~\cite{hetero-hpvm-hdc,gpu-hdtorch,gpu-openhd,fpga-imani-revisiting-hdc-fpga,pim-hdnn-rosing,survey-hdc-edge-intel-progress}. CPU and GPU frameworks exploit wide SIMD units and multi-core parallelism, but they are typically optimized for large batches and struggle to deliver ultra-low latency when processing a continuous stream of small images in real time. FPGAs, in contrast, can implement custom, streaming dataflow architectures that fuse binding, bundling, permutation, and similarity search into deeply pipelined engines with fine-grained control over memory bandwidth and on-chip parallelism~\cite{fpga-imani-revisiting-hdc-fpga}. Yet prior FPGA designs generally target 1D or tabular workloads and do not co-design the HDC encoder and hardware around image patches and hypervector structure.

In this work, we bridge these gaps by co-designing a patch-based HDC image encoder and a primitive-driven FPGA accelerator for real-time image classification. Our key contributions are:

\begin{itemize}
    \item We propose a spatially aware, patch-based HDC encoder that maps local image regions to hypervectors using position- and intensity-dependent banks, aggregates them via permutation and bundling, and refines class hypervectors using similarity-guided online updates. This design achieves higher accuracy than prior HDC image encoders on MNIST and Fashion-MNIST while preserving simple, composable primitives.
    \item We design an end-to-end FPGA accelerator that realizes the complete HDC inference pipeline---patch encoding, global accumulation, and similarity search---as a streaming dataflow architecture. The accelerator exploits parallelism both along the hypervector dimension and across image patches using a patch processor array, a global adder tree, and a similarity engine mapped to an Alveo U280 FPGA.
    \item We provide a comprehensive evaluation across CPU, GPU, and FPGA platforms, showing that our FPGA design delivers orders-of-magnitude lower single-image latency and higher throughput than optimized PyTorch baselines, while maintaining competitive accuracy. Ablation studies on patch size and hypervector dimension further quantify the trade-offs between accuracy and hardware efficiency.
\end{itemize}

\section{Related Work}
\label{sec:related}

\textbf{HDC fundamentals and learning frameworks.}
Hyperdimensional Computing (HDC) has been widely studied as a robust, compact alternative to conventional machine learning, grounded in high-dimensional random representations and simple algebraic primitives~\cite{kanerva,survey-two-parter-part-I,survey-two-parter-part-II}. A variety of methods improve HDC’s encoding and learning capabilities through quantization~\cite{quanthd}, stochastic and memory-centric training~\cite{searchhd}, federated or dynamic updates~\cite{fl-hdc,td-hdc}, and trainable encoders~\cite{trainableHD,trainable-hd-old-version}. Recent work such as LaplaceHDC~\cite{laplacehdc} refines the geometric interpretation of binary hypervectors, while LeHDC~\cite{lehdc} introduces learning-based prototype construction for improved accuracy. These methods demonstrate that modest learning capacity can substantially enhance HDC performance while retaining lightweight, interpretable operations.

\textbf{HDC for image classification.}
Although numerous HDC encoders exist for biosignals, text, and tabular data~\cite{trad-hdc-1,trad-hdc-2,trad-hdc-5-imani-hierarchical-hd}, image-specific HDC remains comparatively limited. Prior work has proposed binarized or static image encodings~\cite{adv-encoding-2-encoding-binarized-img}, descriptor aggregation frameworks~\cite{static-encoding-9-hdc-framework-image-descriptors-cvpr}, and privacy-preserving image representations~\cite{rosing-vision-hd}, but these approaches typically lack explicit spatial structure, patch semantics, or principled permutation-based aggregation. More advanced methods such as ManiHD~\cite{multi-mani-hd-imani} and QuantHD~\cite{quanthd} introduce trainable or quantized pipelines but do not target spatially coherent image encoding comparable to convolutional receptive fields.

\textbf{Hardware acceleration of HDC.}
HDC has been accelerated on CPUs, GPUs, FPGAs, and PIM architectures using vectorization, parallel training, or custom logic~\cite{gpu-hdtorch,gpu-openhd,fpga-imani-revisiting-hdc-fpga,pim-hdnn-rosing,hetero-hpvm-hdc}. GPU-centric designs typically rely on large batches to exploit data parallelism via SIMT, while FPGA implementations leverage dataflow pipelines for low-latency streaming inference. Existing FPGA accelerators, however, primarily target 1D or tabular HDC workloads rather than spatially structured image encoders \cite{searchhd, lehdc, fl-hdc, td-hdc, laplacehdc}. Our work differs by co-designing a patch-based HDC algorithm with an FPGA-friendly architecture, enabling fine-grained patch-level and hypervector-level parallelism tailored for real-time image classification.

\section{Preliminaries}
\label{sec:prelim}

\subsection{Hyperdimensional Computing}

Hyperdimensional Computing (HDC) represents data using high-dimensional
\emph{hypervectors} (HVs) of dimension $D$, typically $D \!\sim\! 10^{3}$–$10^{5}$.  
We use bipolar HVs $\mathbf{h} \in \{-1,+1\}^{D}$, which enable simple,
highly parallel element-wise operations. HDC relies on three fundamental
primitives:

\begin{itemize}
    \item \textbf{Bundling} ($\oplus$): element-wise addition followed by
    bipolarization,  
    $\mathbf{h} = \mathrm{sign}(\mathbf{h}_1 + \mathbf{h}_2 + \cdots)$,  
    producing an HV similar to its constituents.

    \item \textbf{Binding} ($\odot$): element-wise multiplication,  
    $\mathbf{h} = \mathbf{h}_1 \odot \mathbf{h}_2$,  
    generating an HV dissimilar to both inputs and used to associate features
    with positions or symbols.

    \item \textbf{Permutation} ($\pi$): cyclic rotation,  
    $\pi^{t}(\mathbf{h})$,  
    used to encode structural, temporal, or positional information such as a
    sequence index or patch ID.
\end{itemize}

These operations map raw features into high-dimensional representations that
can be compared efficiently using cosine or Hamming similarity. High
dimensionality provides inherent robustness: perturbing a small fraction of
dimensions rarely alters similarity relationships between HVs.

\subsection{HDC Classification Model}
\label{subsec:hdc_classification}

In a standard HDC classifier, each class $c$ is represented by a prototype
hypervector $\mathbf{C}_c \in \{-1,+1\}^{D}$. During training, prototypes are
formed by bundling the encoded HVs of all samples with label $c$:
$\mathbf{C}_c \leftarrow \mathrm{sign}\!\big(\sum_{n : y^{(n)} = c}
\mathbf{H}^{(n)}\big)$. At inference time, an input is encoded into a query HV $\mathbf{H}$ using a
task-specific encoder built from binding, bundling, and permutation.
Classification is performed via nearest-prototype matching:
$\hat{y} = \arg\max_{c}\,\mathrm{sim}(\mathbf{H}, \mathbf{C}_c)$, where
$\mathrm{sim}(\cdot,\cdot)$ denotes cosine similarity or Hamming distance.

\section{Method}
\label{sec:method}
We now describe our primitive-driven HDC algorithm and accelerator for real-time image classification.
Section~\ref{subsec:method_all} presents the patch-based image encoder and
online hypervector learning procedure \cite{onlinehd}, while
Section~\ref{subsec:accelerator} details the FPGA accelerator architecture
that implements this algorithm as a deeply pipelined dataflow engine.
\begin{figure*}
  \centering
  \includegraphics[width=\linewidth]{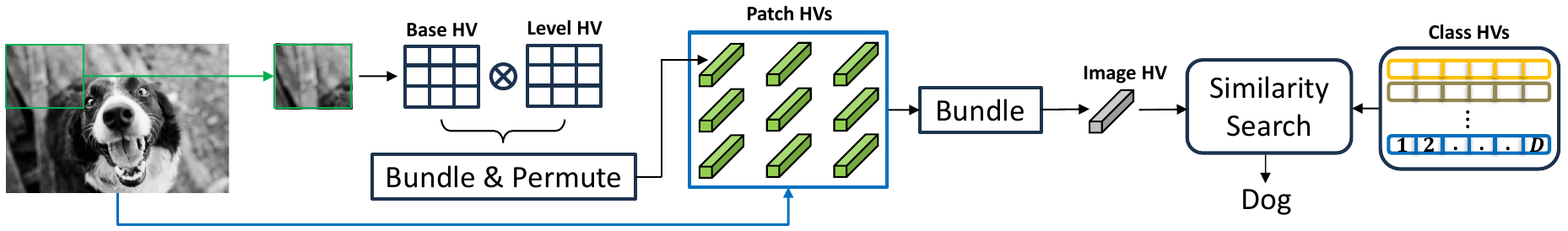}
  \caption{Overall patch-based HDC encoding pipeline (during inference).}
  \label{fig:overview_hdc}
\end{figure*}
\subsection{Image Encoding and Online Hypervector Learning}
\label{subsec:method_all}

Our approach maps each input image $\mathbf{X} \in \mathbb{R}^{H \times W}$ into a
single bipolar hypervector $\mathbf{H} \in \{-1,+1\}^D$, which is then used for
incremental hypervector-based learning.  
The method consists of three stages: pixel-level encoding, patch aggregation with spatial permutation, and global bundling followed by online learning.

\noindent \textbf{Hypervector Banks.}
We initialize two banks of real-valued hypervectors:  
a base bank $\mathbf{B}_{i,j} \in \mathbb{R}^D$ for each pixel position $(i,j)$ and  
a level bank $\mathbf{L}_\ell \in \mathbb{R}^D$ for each quantized intensity level  
$\ell \in \{0,\dots,L-1\}$.  
All hypervectors are sampled i.i.d.\ from a Gaussian distribution and
$\ell_2$–normalized along their dimensional axis,
$\mathbf{B}_{i,j} \leftarrow \mathbf{B}_{i,j} / \|\mathbf{B}_{i,j}\|_2$ and  
$\mathbf{L}_\ell \leftarrow \mathbf{L}_\ell / \|\mathbf{L}_\ell\|_2$.

\noindent \textbf{Pixel Encoding.}
Each pixel value $x_{i,j}$ is first linearly quantized using a scale--zero-point
transformation. Let $s$ and $z$ denote the quantization scale and zero-point that
map the input range of an image to the integer interval $[0,255]$. The
quantized level is
\[
\ell_{i,j} = \mathrm{clip}\!\left( \left\lfloor \frac{x_{i,j}}{s} + z \right\rfloor ,\, 0,\, L-1 \right),
\quad L = 256.
\]
A pixel hypervector is then formed by binding the corresponding base and level
hypervectors:
\[
\mathbf{p}_{i,j} = \mathbf{B}_{i,j} \odot \mathbf{L}_{\ell_{i,j}}.
\]

\noindent \textbf{Patch Aggregation.}
We extract local image regions using square patches of size $M \times M$,
analogous to a convolutional kernel of size $M$.  The patch window is moved
across the image with a chosen stride $r$, producing a grid of
$K_H \times K_W$ patches.  Let $\mathcal{P}_t$ denote the set of pixel
coordinates in the $t$-th patch, where $t \in \{0,\dots,K_H K_W - 1\}$.
For each patch, we sum its pixel hypervectors,
\[
\tilde{\mathbf{h}}_t
=
\sum_{(i,j)\in\mathcal{P}_t} \mathbf{p}_{i,j},
\]
and apply a permutation (cyclic rotation) indexed by the patch ID $t$,
\[
\mathbf{h}_t = \pi^t(\tilde{\mathbf{h}}_t),
\qquad
[\pi^t(\tilde{\mathbf{h}}_t)]_d
=
\tilde{h}_{t,\,(d - t)\bmod D}.
\]

\noindent \textbf{Global Image HV.}
Patch hypervectors are bundled by summation  
$\tilde{\mathbf{H}} = \sum_t \mathbf{h}_t$.  
The final image representation is obtained via bipolar binarization,  
$H_d = +1$ if $\tilde{H}_d \ge 0$ and $H_d = -1$ otherwise,
resulting in $\mathbf{H} \in \{-1,+1\}^D$.

\noindent \textbf{Similarity \& Prediction.}
For each class $c \in \{1,\dots,C\}$ we maintain a class hypervector
$\mathbf{C}_c \in \{-1,+1\}^D$ obtained by bundling the image hypervectors
belonging to that class.  During training, each encoded sample $\mathbf{H}$
is added to the corresponding class sum, and the final class hypervectors
are obtained by applying a bipolarization step,
\[
\mathbf{C}_c \leftarrow \mathrm{sign}\!\left(\sum_{\mathbf{H}\in\mathcal{D}_c}
\mathbf{H}\right),\qquad c=1,\dots,C.
\]

Given an encoded image hypervector $\mathbf{H}$, we classify it by computing
its cosine similarity to each class hypervector,
\[
s_c = \frac{\mathbf{H}^{\top}\mathbf{C}_c}{D}, \qquad s_c \in [-1,+1],
\]
where division by $D$ normalizes the bipolar inner product and makes the score
equivalent to cosine similarity for $\{-1,+1\}$ hypervectors. We get the final predicted label via $\hat{y} = \arg\max_{c} s_c$.

\noindent \textbf{Online Learning.}
To refine the class hypervectors beyond the initial bundled prototypes, we use
a similarity-weighted corrective update applied only to misclassified samples.
Let $s_c \in [-1,+1]$ denote the cosine similarity between an image hypervector
$\mathbf{H}$ and class hypervector $\mathbf{C}_c$, as defined above.  For update
purposes, this score is shifted to the interval $[0,1]$ via $\sigma_c = \frac{s_c + 1}{2}$.
After the initial single-pass construction of the class hypervectors, we run a
few retraining epochs in which each misclassified sample with true label $y$
and predicted label $\hat{y}$ triggers the updates
\[
\mathbf{C}_y      \leftarrow \mathbf{C}_y      + \eta (1 - \sigma_y)\mathbf{H},
\qquad
\mathbf{C}_{\hat{y}} \leftarrow \mathbf{C}_{\hat{y}} - \eta (1 - \sigma_{\hat{y}})\mathbf{H},
\]
where $\eta$ is a learning rate.  After retraining, the class hypervectors are
binarized using $\mathrm{sign}(\cdot)$ and the resulting bipolar vectors are
used for inference.

\noindent \textbf{Overall Algorithm.}
Algorithm~\ref{alg:patch_onlinehd} summarizes the full pipeline, with the inference procedure visualized in Fig. \ref{fig:overview_hdc}.

\begin{algorithm}[t]
  \caption{Patch-based bipolar hypervector encoding and online learning}
  \label{alg:patch_onlinehd}
  \begin{algorithmic}[1]
    \State \textbf{Input:} training set $\{(\mathbf{X}^{(n)}, y^{(n)})\}$, 
           test set $\{\mathbf{X}^{(m)}\}$, dimension $D$, patch size $M$, 
           stride $r$, levels $L$, quantization scale $s$, zero-point $z$,
           learning rate $\eta$
    \State Initialize hypervector banks $\mathbf{B}_{i,j}$ and $\mathbf{L}_{\ell}$
    \State Initialize class hypervectors $\mathbf{C}_c \gets \mathbf{0}$ for all $c$

    \State // image encoding
    \Function{EncodeImage}{$\mathbf{X}$}
      \State Quantize pixels: $\ell_{i,j} \gets \mathrm{clip}\!\left(\left\lfloor X_{i,j}/s + z \right\rfloor, 0, L-1\right)$
      \State $\mathbf{p}_{i,j} \gets \mathbf{B}_{i,j} \odot \mathbf{L}_{\ell_{i,j}}$
      \For{each patch $t$ with pixels $\mathcal{P}_t$}
        \State $\tilde{\mathbf{h}}_t \gets \sum_{(i,j)\in\mathcal{P}_t} \mathbf{p}_{i,j}$
        \State $\mathbf{h}_t \gets \pi^t(\tilde{\mathbf{h}}_t)$
      \EndFor
      \State $\tilde{\mathbf{H}} \gets \sum_t \mathbf{h}_t$
      \State $\mathbf{H} \gets \mathrm{sign}(\tilde{\mathbf{H}})$ \Comment{bipolar binarization}
      \State \Return $\mathbf{H}$
    \EndFunction

    \State // online learning (retraining over misclassified samples)
    \For{each $(\mathbf{X}^{(n)}, y^{(n)})$ in training set}
      \State $\mathbf{H} \gets \Call{EncodeImage}{\mathbf{X}^{(n)}}$
      \State Compute cosine similarities: $s_c \gets \mathbf{H}^\top \mathbf{C}_c / D$ for all $c$
      \State Predicted label: $\hat{y} \gets \arg\max_c s_c$
      \If{$\hat{y} \neq y^{(n)}$}
        \State $\sigma_{y^{(n)}}      \gets (s_{y^{(n)}} + 1)/2$
        \State $\sigma_{\hat{y}} \gets (s_{\hat{y}} + 1)/2$
        \State $\mathbf{C}_{y^{(n)}} \gets \mathbf{C}_{y^{(n)}} + \eta (1 - \sigma_{y^{(n)}})\mathbf{H}$
        \State $\mathbf{C}_{\hat{y}} \gets \mathbf{C}_{\hat{y}} - \eta (1 - \sigma_{\hat{y}})\mathbf{H}$
      \EndIf
    \EndFor

    \State // final binarization
    \State $\mathbf{C}_c \gets \mathrm{sign}(\mathbf{C}_c)$ for all $c$

    \State // inference
    \For{each test image $\mathbf{X}^{(m)}$}
      \State $\mathbf{H} \gets \Call{EncodeImage}{\mathbf{X}^{(m)}}$
      \State $\hat{y}^{(m)} \gets \arg\max_c \mathbf{H}^{\top}\mathbf{C}_c$
    \EndFor

  \end{algorithmic}
\end{algorithm}

\subsection{Accelerator Architecture}
\label{subsec:accelerator}

\begin{figure*}[t]
    \centering
    \begin{minipage}{0.49\textwidth}
        \centering
        \includegraphics[width=\linewidth]{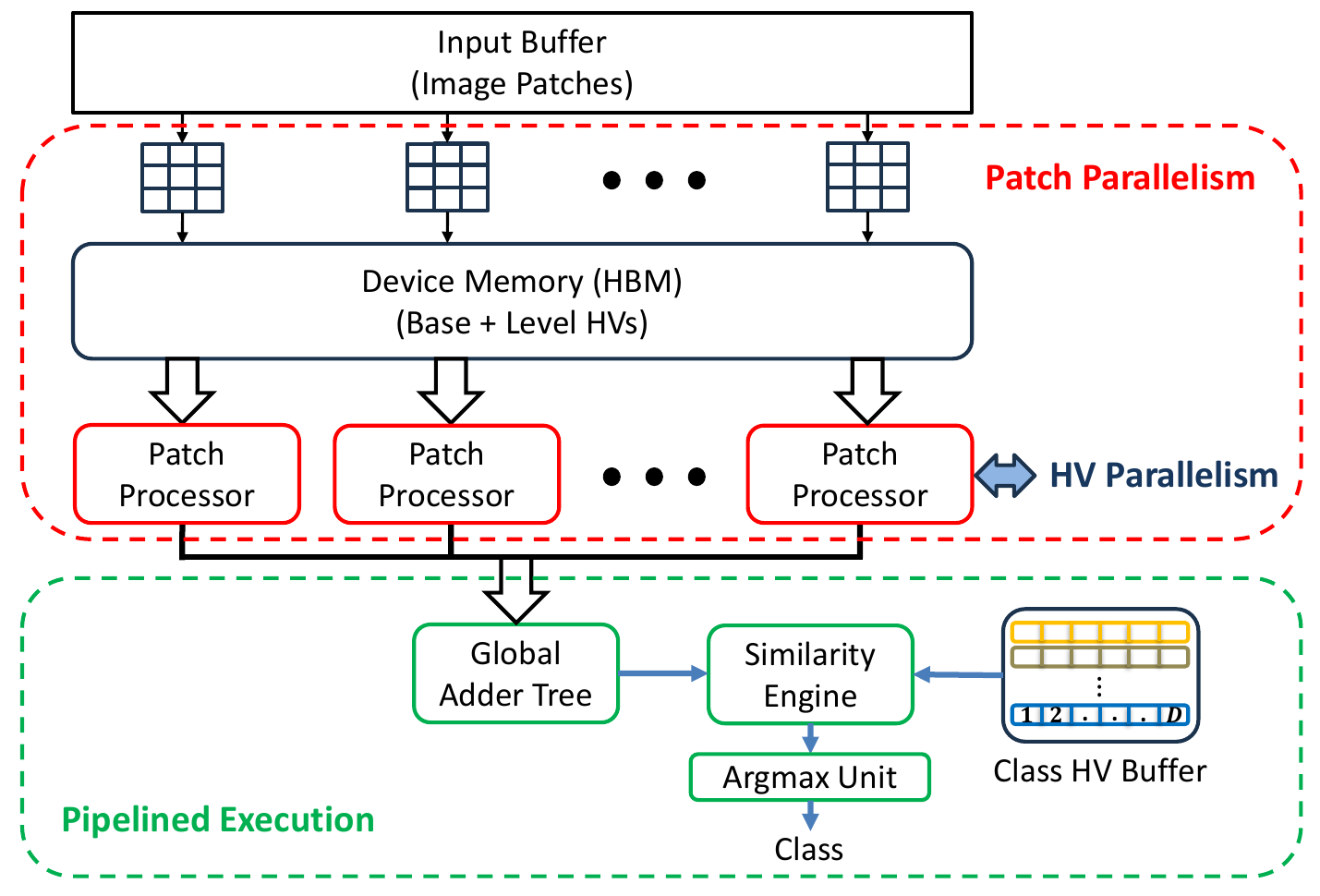}\\
        \small\textbf{(a) Accelerator Microarchitecture}
    \end{minipage}
    \hfill
    \begin{minipage}{0.47\textwidth}
        \centering
        \includegraphics[width=\linewidth]{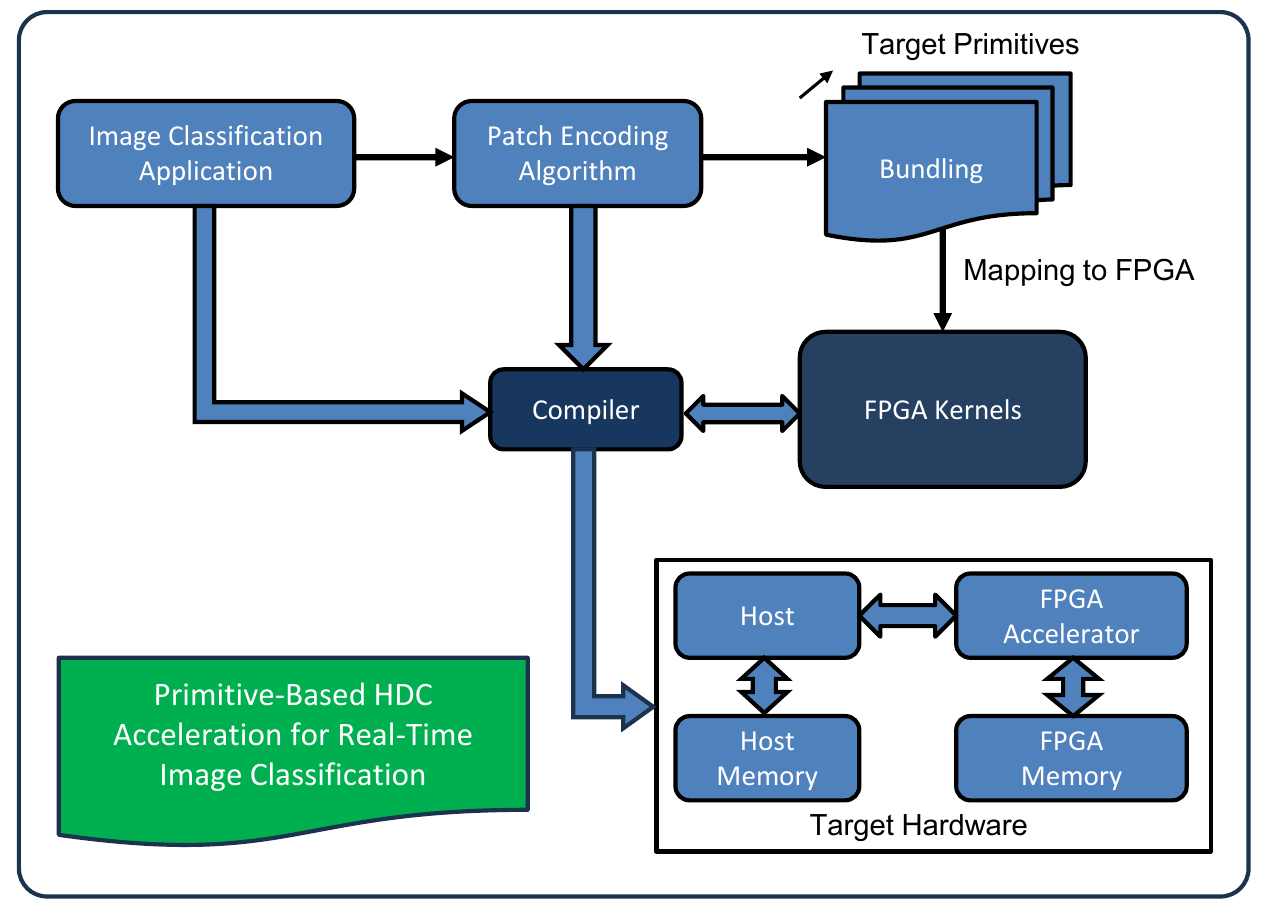}\\
        \small\textbf{(b) Mapping to FPGA Kernels}
    \end{minipage}

    \vspace{1mm}
    \caption{
        Overview of the proposed primitive-driven HDC acceleration framework.
        (a) The accelerator architecture exposes both patch-level and hypervector-level
        parallelism using multiple patch processors, a global adder tree, a similarity
        engine, and a lightweight argmax unit.
        (b) End-to-end system mapping from the patch-based encoding algorithm
        to FPGA compute kernels and host–device execution flow.
    }
    \label{fig:hdc_accel_overview}
\end{figure*}

Our accelerator implements the complete patch-based HDC inference as a deeply
pipelined dataflow architecture composed of four compute engines: a patch
processor array, a global adder tree, a similarity engine, and a lightweight
argmax unit. Parallelism is exploited across both the hypervector
dimensionality and the total number of image patches. 

\noindent \textbf{Patch Processor Array.}
The accelerator instantiates $P_{\mathrm{patch}}$ parallel patch processors,
each responsible for encoding one $M{\times}M$ image patch.  
A patch processor contains $P_D$ parallel vector (MAC) lanes that operate across the
hypervector dimension. For each patch, the processor streams in the $M^2$ base
and level hypervectors, performs element-wise binding using the $P_D$ lanes,
and accumulates the results into a local $P_D$-wide buffer. After processing
all $M^2$ pixels, the processor applies the patch-dependent permutation via a
barrel-shifter and emits a partial patch hypervector of length $P_D$. This engine forms the backbone of the accelerator, enabling
parallelism both across patches and across hypervector dimensions.

\noindent \textbf{Global Adder Tree.}
Patch hypervectors produced by the processor array are streamed into a global
adder tree. The tree reduces the $P_{\mathrm{patch}}$ incoming patch outputs into
a single partial image hypervector corresponding to the current $P_D$-wide
segment of the global $D$-dimensional representation.  
The adder tree is fully pipelined and processes one set of patch results per
cycle.


\noindent \textbf{Similarity Engine.}
The similarity engine computes dot-product similarity between each incoming
hypervector segment and all class hypervectors. It consists of a bank of
$P_D$ parallel multiply--accumulate (MAC) units that process one
$P_D$-dimensional segment at a time. For each segment, the engine multiplies
the $P_D$ elements of the image hypervector with the corresponding
$P_D$ elements of every class hypervector, accumulating the partial products
into a score buffer whose size equals the number of classes. As successive
segments are streamed in, the accumulated scores are updated.
After all $\lceil D / P_D \rceil$ segments have been processed, the score
buffer contains the complete similarity values for all classes.

\noindent \textbf{Argmax Unit.}
Once all hypervector segments have been processed, the accumulated
similarity scores for all classes are streamed into a compact argmax unit.  
This unit identifies the predicted class label using a pipelined comparator
tree.

\noindent \textbf{End-to-End Streaming Dataflow.}
The accelerator operates as a multi-stage streaming pipeline (Fig. \ref{fig:hdc_accel_overview}).  
Base and level hypervectors are fetched from on-device memory and delivered to the $P_{\mathrm{patch}}$ patch processors, each of which encodes its assigned $M{\times}M$ patch using $P_D$ parallel multiply--accumulate lanes. As soon as a partial patch hypervector segment is generated, it is forwarded to the global adder tree, which progressively accumulates contributions from all patches to form the corresponding segment of the image hypervector. The similarity engine consumes each segment immediately, performing dot-product
similarity against all class hypervectors and updating a running similarity
buffer. After all hypervector segments have been processed, the accumulated
scores are passed to a lightweight argmax unit to determine the final predicted class. This design exposes two complementary forms of parallelism:  
(i) \emph{hypervector parallelism} via $P_D$ concurrent MAC operations per
cycle per patch processor, and  
(ii) \emph{patch-level parallelism} via $P_{\mathrm{patch}}$ independent patch
processors.  
Together, these dimensions enable a fully pipelined, low-latency realization of
the HDC inference pipeline in Algorithm~\ref{alg:patch_onlinehd}.

\section{Experiments}
\label{sec:experiments}

\subsection{Datasets}

We evaluate our method on two standard image-classification benchmarks:
MNIST \cite{mnist} and Fashion-MNIST \cite{fashion-mnist}. Both datasets consist of $28{\times}28$ grayscale
images across 10 classes. Following common practice in HDC-based image
encoding, all images are zero-padded to $32{\times}32$ before patch extraction
and quantized to $[0,255]$ using an affine scale--zero-point mapping.

\begin{table}
    \centering
    \caption{Summary of datasets used in evaluation.}
    \label{tab:datasets}
    \begin{tabular}{lccc}
        \toprule
        Dataset & Image Size & \# Classes & Train / Test \\
        \midrule
        MNIST & $28{\times}28$ & 10 & 60K / 10K \\
        Fashion-MNIST & $28{\times}28$ & 10 & 60K / 10K \\
        \bottomrule
    \end{tabular}
\end{table}

\subsection{Implementation Setup}

\begin{table}
\centering
\caption{Hardware platforms used in our evaluation.}
\label{tab:platforms}
\vspace{1mm}

\resizebox{\columnwidth}{!}{
\begin{tabular}{lccc}
\toprule
\textbf{Platforms} & \textbf{CPU} & \textbf{GPU} & \textbf{FPGA} \\
\midrule

Device &
EPYC 7313 (2×) &
RTX 6000 Ada &
Alveo U280 \\

Process Node &
TSMC 7\,nm &
TSMC 4N (4\,nm) &
TSMC 16\,nm \\

Compute Units &
32C/64T &
18{,}176 CUDA cores &
9024 DSP slices \\

Memory &
DDR4 (Host) &
48\,GB GDDR6 &
8\,GB HBM2 \\

Bandwidth &
205\,GB/s &
960\,GB/s &
460\,GB/s \\

Frequency &
3.7\,GHz &
2.5\,GHz (boost) &
300--600\,MHz \\

Peak FP32 &
-- &
91.1\,TFLOPS &
-- \\
\bottomrule
\end{tabular}
}
\end{table}

\noindent \textbf{Hardware Platforms.}
We evaluate our approach on three representative platforms summarized in
Table~\ref{tab:platforms}. 
For the CPU baseline, we use a dual-socket AMD EPYC~7313 server with
32~cores / 64~threads and DDR4 host memory. 
The GPU baseline runs on an NVIDIA RTX~6000 Ada accelerator with 18{,}176 CUDA
cores and 48\,GB of GDDR6 memory, providing up to 960\,GB/s of peak memory
bandwidth. 
For FPGA-based acceleration, we implement our HDC architecture on an AMD
Alveo U280 data-center FPGA card equipped with 9024 DSP slices, on-board HBM2
memory (460\,GB/s aggregate bandwidth), and a high-speed PCIe host interface.

\noindent \textbf{FPGA Implementation.}
The proposed accelerator (Section~\ref{sec:method}) is implemented in C/C++ using AMD Vitis High-Level Synthesis (HLS) and integrated using the AMD Vitis Unified IDE~v2024.2 toolchain. 
We target a core clock frequency of 250\,MHz on the U280 and fix the
hypervector-parallelism and patch-parallelism parameters to 
$P_{\mathrm{D}} = 256$ and $P_{\mathrm{patch}} = 16$, respectively. 
These choices provide a practical balance between exploiting parallelism along
the hypervector dimension and across image patches, while remaining within the
DSP, BRAM, and HBM bandwidth constraints of the device. 

\noindent \textbf{Software Stack.}
We implement the patch-based HDC algorithm and online hypervector learning
pipeline in Python using PyTorch for both CPU and GPU execution. 
CPU baselines are run with the number of worker threads set equal to the
available physical cores on the EPYC server, while GPU baselines leverage
CUDA and vendor-optimized libraries (cuBLAS/cuDNN) through PyTorch’s default
backend. 
The same PyTorch implementation is used to report accuracy on MNIST and
Fashion-MNIST and to profile end-to-end latency and throughput on the CPU and
GPU platforms.

\subsection{Hypervector Encoding Parameters}

All experiments use a hypervector dimension of $D=10{,}000$ and 
$256$ quantization levels. We adopt non-overlapping $3{\times}3$ patches with stride~$3$ on $32{\times}32$ MNIST and Fashion-MNIST inputs, resulting in $K_H = K_W = 10$ and a total of $100$ patches per image.

\subsection{Accuracy Evaluation}

\begin{table}
\centering
\caption{MNIST classification accuracy comparison with standard HDC baselines.}
\label{tab:mnist_accuracy}
\vspace{1mm}
\begin{tabular}{lcc}
\toprule
\textbf{Method} & \textbf{Accuracy (\%)} & \textbf{Dimension} \\
\midrule
SearcHD \cite{searchhd}      & 84.43 & 10,000 \\
FL-HDC \cite{fl-hdc}         & 88.00 & 10,000 \\
TD-HDC \cite{td-hdc}         & 88.92 & 5,000  \\
QuantHD \cite{quanthd}       & 89.28 & 10,000 \\
LeHDC \cite{lehdc}           & 94.74 & 10,000 \\
LaplaceHDC \cite{laplacehdc} & 94.59 & 10,000 \\
\midrule
\textbf{Ours}                & \textbf{95.67} & \textbf{10,000} \\
\bottomrule
\end{tabular}
\end{table}

\noindent \textbf{MNIST.}
Our patch-based HDC encoder achieves an accuracy of \textbf{95.67\%} on MNIST
using a $10{,}000$-dimensional image hypervector. This surpasses prior HDC
baselines—including SearcHD~\cite{searchhd}, FL-HDC~\cite{fl-hdc},
TD-HDC~\cite{td-hdc}, and QuantHD~\cite{quanthd}, which report accuracies in
the $84$–$89\%$ range—and also exceeds more advanced binary or quantized
architectures such as LeHDC~\cite{lehdc} and LaplaceHDC~\cite{laplacehdc},
both around $94.6$–$94.7\%$.
These results show that a simple patch-aware HDC encoder can rival or surpass
approaches that rely on heavy learning procedures~\cite{trainableHD} or
computationally intensive manifold-based representations~\cite{multi-mani-hd-imani}.

\begin{table}
\centering
\caption{Fashion-MNIST accuracy comparison.}
\label{tab:fashion_accuracy}
\vspace{1mm}
\begin{tabular}{lcc}
\toprule
\textbf{Method} & \textbf{Accuracy (\%)} & \textbf{Dimension} \\
\midrule
LaplaceHDC \cite{laplacehdc} & 83.51 & 10,000 \\
\midrule
\textbf{Ours}                & \textbf{85.14} & \textbf{10,000} \\
\bottomrule
\end{tabular}
\end{table}

\noindent \textbf{Fashion-MNIST.}
On Fashion-MNIST, our encoder achieves \textbf{85.14\%} accuracy using
$10{,}000$-dimensional hypervectors, exceeding the reported $83.51\%$ of
LaplaceHDC~\cite{laplacehdc}. This demonstrates that a lightweight,
patch-based HDC encoder can remain competitive on more challenging visual
datasets without relying on additional optimization stages.

\subsection{Latency and Throughput Evaluation}
\label{subsec:latency_throughput}

To assess end-to-end inference performance, we compare our FPGA accelerator
against optimized CPU (PyTorch, multi-threaded) and GPU (PyTorch CUDA)
baselines. Latency corresponds to single-image inference (batch size 1), and
throughput is measured as images per second (batch size 8 for CPU/GPU).
For FPGA, throughput is simply the inverse of single-image latency, as the
accelerator processes one image at a time.

\begin{figure}[t]
    \centering
    \begin{minipage}{0.48\columnwidth}
        \centering
        \includegraphics[width=\linewidth]{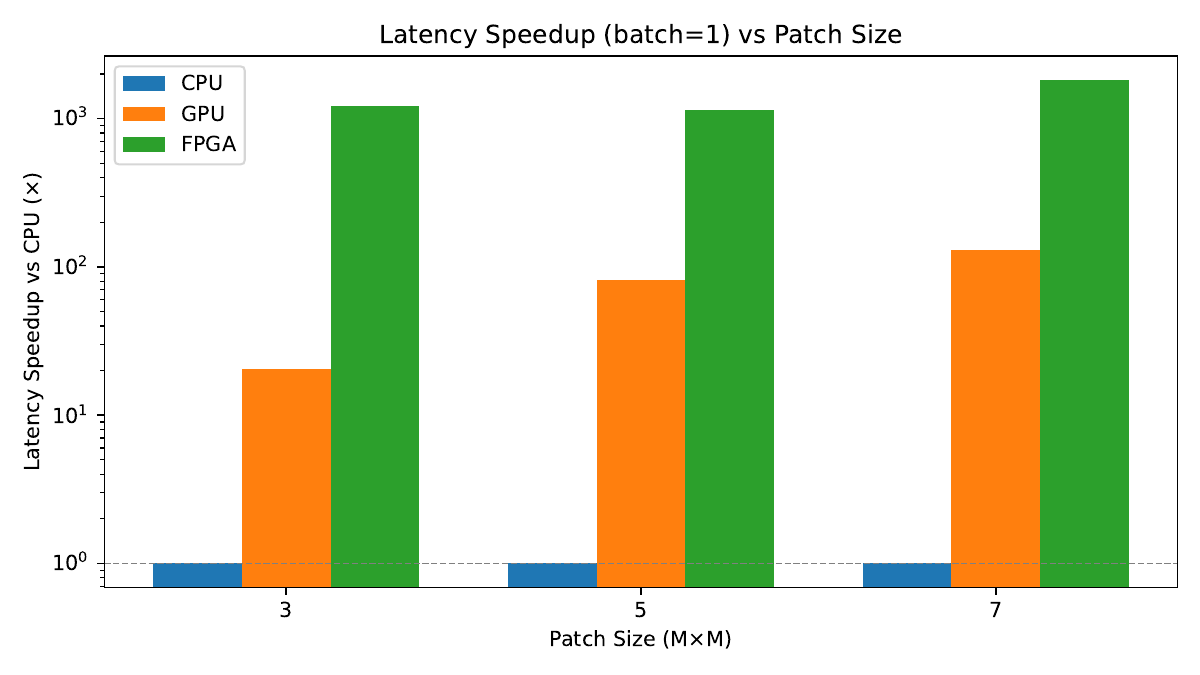}
    \end{minipage}\hfill
    \begin{minipage}{0.48\columnwidth}
        \centering
        \includegraphics[width=\linewidth]{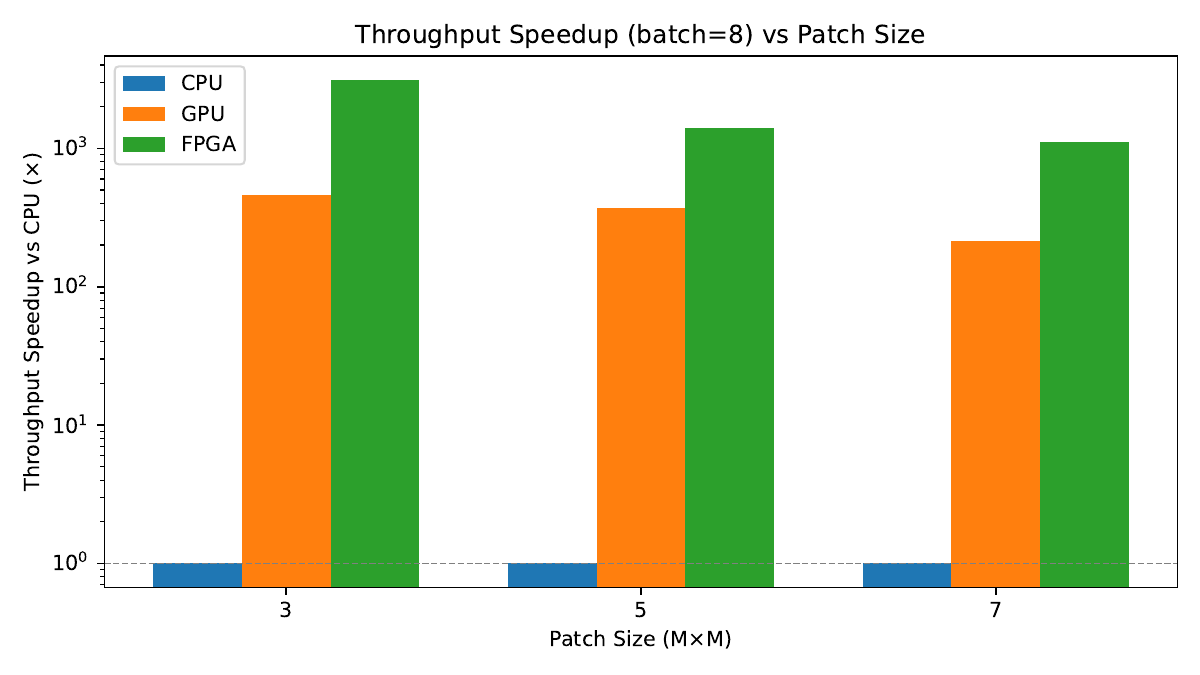}
    \end{minipage}

    \vspace{2mm}

    \begin{minipage}{0.48\columnwidth}
        \centering
        \includegraphics[width=\linewidth]{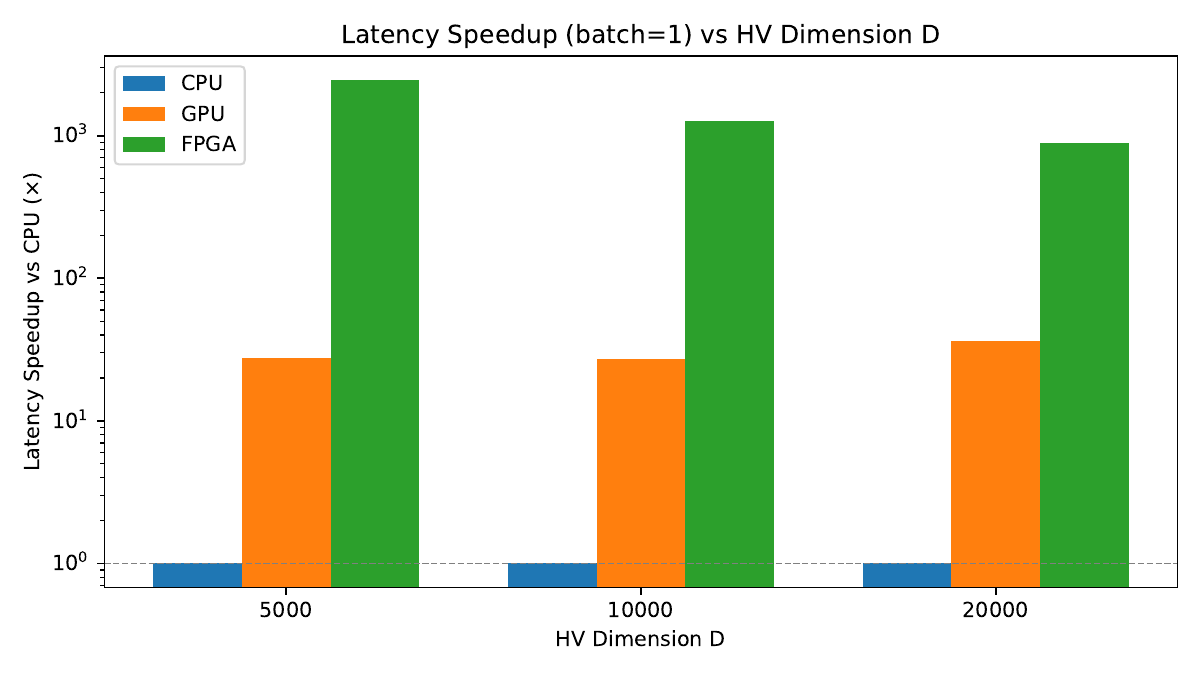}
    \end{minipage}\hfill
    \begin{minipage}{0.48\columnwidth}
        \centering
        \includegraphics[width=\linewidth]{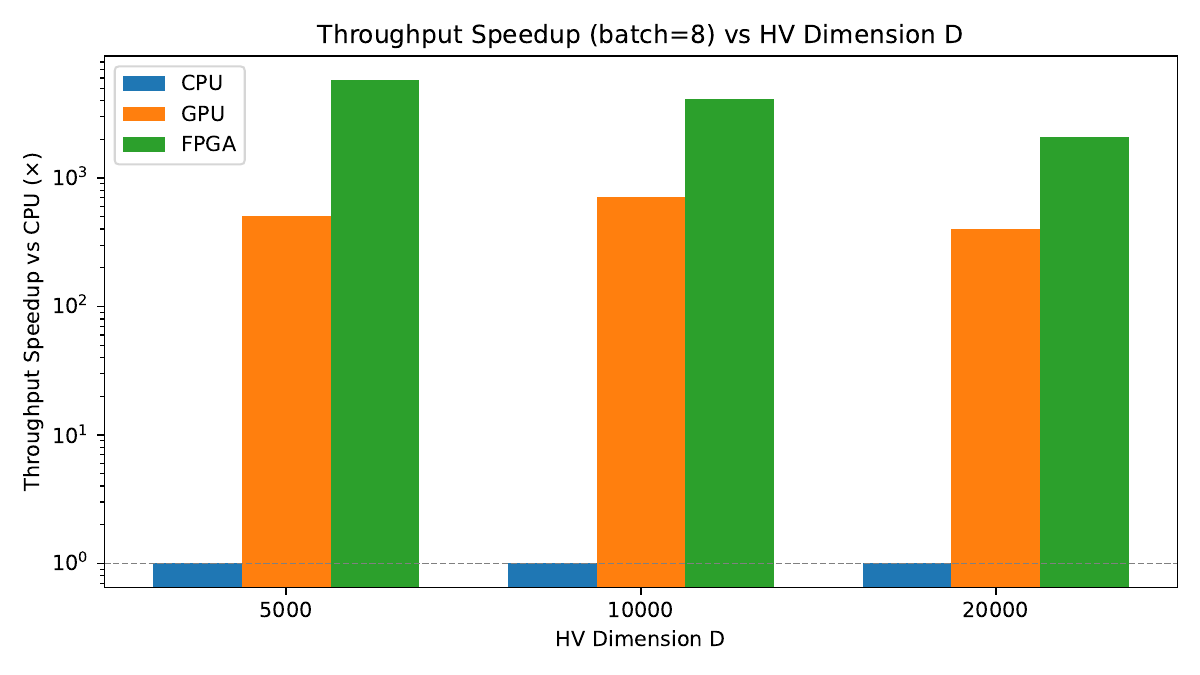}
    \end{minipage}

    \caption{
    Latency and throughput speedup relative to the CPU baseline.
    (Top) Effects of patch size using $D{=}10{,}000$.
    (Bottom) Scaling with hypervector dimension $D$.
    CPU speedup baseline is $1\times$.
    }
    \label{fig:lat_thru_speeds}
\end{figure}

\noindent\textbf{Results.}
Figure~\ref{fig:lat_thru_speeds} summarizes the performance. FPGA achieves
sub-millisecond inference across all tested configurations
(Table~\ref{tab:latency_raw}), consistently outperforming both CPU and GPU.
CPU exhibits high latency due to sequential patch accumulation and repeated
memory transfers, while GPU benefits from tensor-level parallelism but still
incurs kernel-launch overheads and limited utilization at small batch sizes.
FPGA maintains a fixed, deeply pipelined datapath that exploits parallelism
along both the patch and hypervector dimensions, yielding
$50\text{--}60\times$ lower latency than GPU and over $1000\times$ lower
latency than CPU. This improvement arises from mapping the entire HDC
end-to-end workflow—patch encoding, global accumulation, and similarity
search—onto a custom streaming dataflow architecture with dedicated compute
engines and no software-level overheads.

Despite GPUs typically benefiting from larger batch sizes, FPGA achieves higher
throughput even when GPU operates with batch size $B{=}8$. This behavior
reflects the ability of a deeply pipelined streaming accelerator to sustain
near-constant throughput once the pipeline is filled, whereas GPU throughput at
small batches is constrained by launch latency, synchronization, and
underutilization of its wide SIMD fabric. These characteristics make FPGA
particularly attractive for real-time, latency-critical HDC-based image
classification workloads. The results in Table~\ref{tab:latency_raw} show FPGA providing approximately
$1300\times$ lower latency than CPU and up to $60\times$ lower latency than
GPU.

\begin{table}[t]
\centering
\caption{End-to-end MNIST inference latency (batch size = 1).}
\label{tab:latency_raw}
\vspace{1mm}
\begin{tabular}{lcc}
\toprule
\textbf{Platform} & \textbf{Latency (ms)} & \textbf{Throughput (img/s)} \\
\midrule
CPU   & 118.0 & 8.5  \\
GPU   & 4.4   & 1906 \\
FPGA  & \textbf{0.09}  & \textbf{11{,}141} \\
\bottomrule
\end{tabular}
\end{table}

\noindent

\subsection{Accuracy Ablations}
\begin{figure}[t]
    \centering
    \includegraphics[width=\columnwidth]{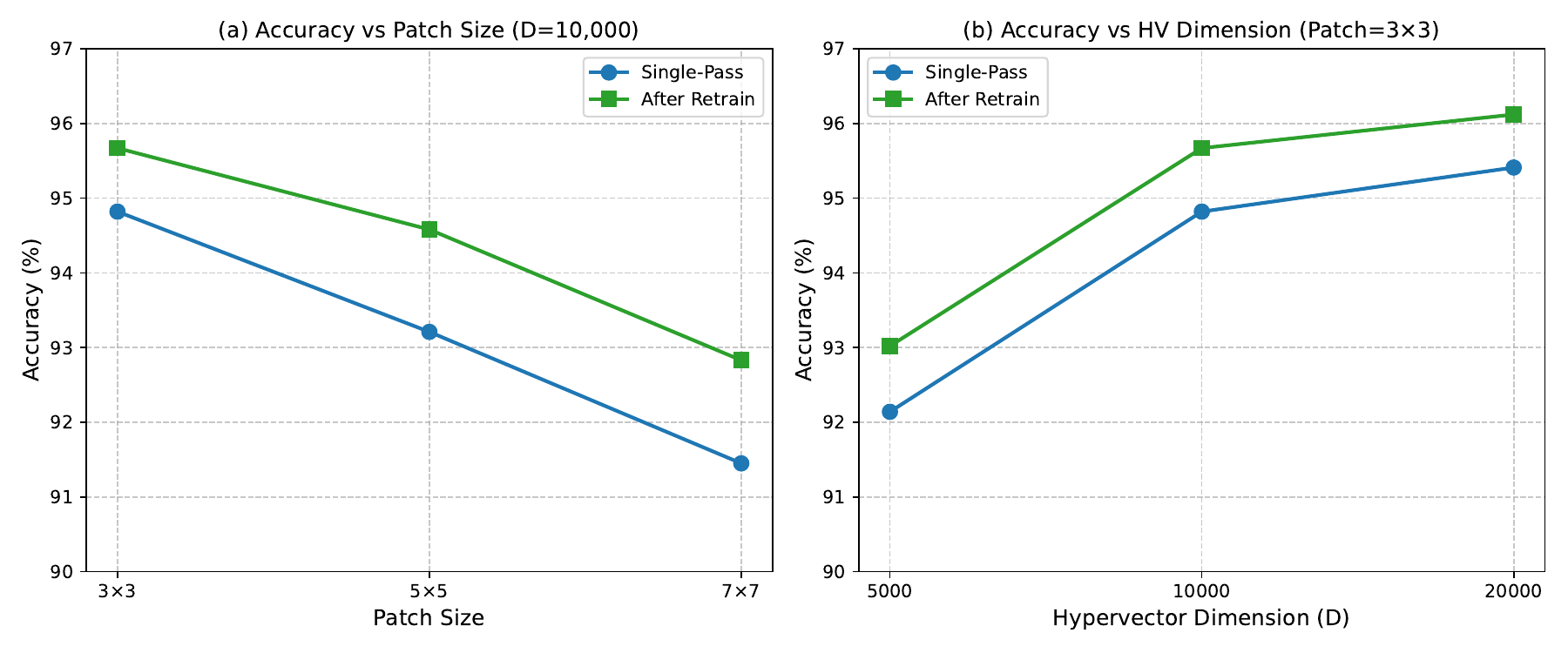}
    \caption{
    Effect of patch size and hypervector dimension on MNIST accuracy.
    Results are shown for both single-pass training and OnlineHD-style
    retraining. Smaller patches preserve discriminative structure, and
    accuracy improves with larger $D$ but saturates beyond $10{,}000$.
    }
    \label{fig:accuracy_ablation}
\end{figure}

We evaluate the impact of patch size and hypervector dimension on MNIST
classification accuracy. Figure~\ref{fig:accuracy_ablation} shows that smaller
patches consistently yield higher accuracy: $3{\times}3$ patches achieve
$95.67\%$, outperforming $5{\times}5$ and $7{\times}7$ by $1$--$3\%$. This trend
is expected, as MNIST digits contain fine-grained strokes that are best captured
with high spatial resolution. Increasing the hypervector dimension $D$ also
improves accuracy, but with diminishing returns: accuracy rises from
$93.02\%$ (for $D=5{,}000$) to $95.67\%$ ($D=10{,}000$), and only marginally to
$96.12\%$ at $D=20{,}000$. Notably, $D{=}10{,}000$ offers the best trade-off
between accuracy and computational cost, which is particularly important for FPGA
deployment.

Retraining via OnlineHD’s similarity-guided updates consistently improves
accuracy by $0.8$--$1.4\%$ across all configurations, confirming that even a small
number of refinement epochs significantly sharpens class hypervectors. Overall,
these ablations validate that the chosen configuration ($3{\times}3$ patches,
$D{=}10{,}000$) provides the best balance of accuracy, efficiency, and hardware
friendliness.

\section{Conclusion}
\label{sec:conclusion}

We presented a spatially aware, patch-based HDC encoder and a primitive-driven FPGA accelerator for real-time image classification. By co-designing the algorithm and architecture around HDC’s binding, bundling, and permutation primitives, our approach achieves competitive accuracy while delivering orders-of-magnitude lower latency than CPU and GPU baselines.

\bibliographystyle{IEEEtran}
\bibliography{main}

@ARTICLE{mnist,
  author={Deng, Li},
  journal={IEEE Signal Processing Magazine}, 
  title={The MNIST Database of Handwritten Digit Images for Machine Learning Research [Best of the Web]}, 
  year={2012},
  volume={29},
  number={6},
  pages={141-142},
  doi={10.1109/MSP.2012.2211477}}

@article{fashion-mnist,
  title={Fashion-mnist: a novel image dataset for benchmarking machine learning algorithms},
  author={Xiao, Han and Rasul, Kashif and Vollgraf, Roland},
  journal={arXiv preprint arXiv:1708.07747},
  year={2017}
}

@inproceedings{lehdc,
  title={Lehdc: Learning-based hyperdimensional computing classifier},
  author={Duan, Shijin and Liu, Yejia and Ren, Shaolei and Xu, Xiaolin},
  booktitle={Proceedings of the 59th ACM/IEEE Design Automation Conference},
  pages={1111--1116},
  year={2022}
}

@article{quanthd,
  title={Quanthd: A quantization framework for hyperdimensional computing},
  author={Imani, Mohsen and Bosch, Samuel and Datta, Sohum and Ramakrishna, Sharadhi and Salamat, Sahand and Rabaey, Jan M and Rosing, Tajana},
  journal={IEEE Transactions on Computer-Aided Design of Integrated Circuits and Systems},
  volume={39},
  number={10},
  pages={2268--2278},
  year={2019},
  publisher={IEEE}
}

@article{searchhd,
  title={Searchd: A memory-centric hyperdimensional computing with stochastic training},
  author={Imani, Mohsen and Yin, Xunzhao and Messerly, John and Gupta, Saransh and Niemier, Michael and Hu, Xiaobo Sharon and Rosing, Tajana},
  journal={IEEE Transactions on Computer-Aided Design of Integrated Circuits and Systems},
  volume={39},
  number={10},
  pages={2422--2433},
  year={2019},
  publisher={IEEE}
}

@inproceedings{fl-hdc,
  title={Fl-hdc: Hyperdimensional computing design for the application of federated learning},
  author={Hsieh, Cheng-Yen and Chuang, Yu-Chuan and Wu, An-Yeu Andy},
  booktitle={2021 IEEE 3rd International Conference on Artificial Intelligence Circuits and Systems (AICAS)},
  pages={1--5},
  year={2021},
  organization={IEEE}
}

@inproceedings{td-hdc,
  title={Dynamic hyperdimensional computing for improving accuracy-energy efficiency trade-offs},
  author={Chuang, Yu-Chuan and Chang, Cheng-Yang and Wu, An-Yeu Andy},
  booktitle={2020 IEEE Workshop on Signal Processing Systems (SiPS)},
  pages={1--5},
  year={2020},
  organization={IEEE}
}

@article{trainableHD,
  title={Advancing Hyperdimensional Computing Based on Trainable Encoding and Adaptive Training for Efficient and Accurate Learning},
  author={Kim, Jiseung and Lee, Hyunsei and Imani, Mohsen and Kim, Yeseong},
  journal={ACM Transactions on Design Automation of Electronic Systems},
  volume={29},
  number={5},
  pages={1--25},
  year={2024},
  publisher={ACM New York, NY}
}

@inproceedings{multi-mani-hd-imani,
  title={Manihd: Efficient hyper-dimensional learning using manifold trainable encoder},
  author={Zou, Zhuowen and Kim, Yeseong and Najafi, M Hassan and Imani, Mohsen},
  booktitle={2021 Design, Automation \& Test in Europe Conference \& Exhibition (DATE)},
  pages={850--855},
  year={2021},
  organization={IEEE}
}

@article{laplacehdc,
  title={Laplace-HDC: Understanding the geometry of binary hyperdimensional computing},
  author={Pourmand, Saeid and Whiting, Wyatt D and Aghasi, Alireza and Marshall, Nicholas F},
  journal={Journal of Artificial Intelligence Research},
  volume={82},
  pages={1293--1323},
  year={2025}
}

@article{survey-hdc-stoch-framework,
  title={Hyperdimensional computing: a framework for stochastic computation and symbolic AI},
  author={Heddes, Mike and Nunes, Igor and Givargis, Tony and Nicolau, Alexandru and Veidenbaum, Alex},
  journal={Journal of Big Data},
  volume={11},
  number={1},
  pages={145},
  year={2024},
  publisher={Springer}
}

@article{survey-two-parter-part-I,
  title={A survey on hyperdimensional computing aka vector symbolic architectures, part i: Models and data transformations},
  author={Kleyko, Denis and Rachkovskij, Dmitri A and Osipov, Evgeny and Rahimi, Abbas},
  journal={ACM Computing Surveys},
  volume={55},
  number={6},
  pages={1--40},
  year={2022},
  publisher={ACM New York, NY}
}

@article{survey-two-parter-part-II,
  title={A survey on hyperdimensional computing aka vector symbolic architectures, part ii: Applications, cognitive models, and challenges},
  author={Kleyko, Denis and Rachkovskij, Dmitri and Osipov, Evgeny and Rahimi, Abbas},
  journal={ACM Computing Surveys},
  volume={55},
  number={9},
  pages={1--52},
  year={2023},
  publisher={ACM New York, NY}
}

@article{kanerva,
  title={Hyperdimensional computing: An introduction to computing in distributed representation with high-dimensional random vectors},
  author={Kanerva, Pentti},
  journal={Cognitive computation},
  volume={1},
  pages={139--159},
  year={2009},
  publisher={Springer}
}

@article{survey-hdc-edge-intel-progress,
  title={Recent progress and development of hyperdimensional computing (hdc) for edge intelligence},
  author={Chang, Cheng-Yang and Chuang, Yu-Chuan and Huang, Chi-Tse and Wu, An-Yeu},
  journal={IEEE Journal on Emerging and Selected Topics in Circuits and Systems},
  volume={13},
  number={1},
  pages={119--136},
  year={2023},
  publisher={IEEE}
}

@article{trad-hdc-1,
  title={Efficient biosignal processing using hyperdimensional computing: Network templates for combined learning and classification of ExG signals},
  author={Rahimi, Abbas and Kanerva, Pentti and Benini, Luca and Rabaey, Jan M},
  journal={Proceedings of the IEEE},
  volume={107},
  number={1},
  pages={123--143},
  year={2018},
  publisher={IEEE}
}

@inproceedings{trad-hdc-2,
  title={A robust and energy-efficient classifier using brain-inspired hyperdimensional computing},
  author={Rahimi, Abbas and Kanerva, Pentti and Rabaey, Jan M},
  booktitle={Proceedings of the 2016 international symposium on low power electronics and design},
  pages={64--69},
  year={2016}
}

@INPROCEEDINGS{trad-hdc-3-imani-voice-hd,
  author={Imani, Mohsen and Kong, Deqian and Rahimi, Abbas and Rosing, Tajana},
  booktitle={2017 IEEE International Conference on Rebooting Computing (ICRC)}, 
  title={VoiceHD: Hyperdimensional Computing for Efficient Speech Recognition}, 
  year={2017},
  volume={},
  number={},
  pages={1-8},
  doi={10.1109/ICRC.2017.8123650}}

@inproceedings{trad-hdc-4-imani-dna,
  title={Hdna: Energy-efficient dna sequencing using hyperdimensional computing},
  author={Imani, Mohsen and Nassar, Tarek and Rahimi, Abbas and Rosing, Tajana},
  booktitle={2018 IEEE EMBS International Conference on Biomedical \& Health Informatics (BHI)},
  pages={271--274},
  year={2018},
  organization={IEEE}
}

@inproceedings{trad-hdc-5-imani-hierarchical-hd,
  title={Hierarchical hyperdimensional computing for energy efficient classification},
  author={Imani, Mohsen and Huang, Chenyu and Kong, Deqian and Rosing, Tajana},
  booktitle={Proceedings of the 55th Annual Design Automation Conference},
  pages={1--6},
  year={2018}
}

@article{static-encoding-1-programmable-hdc,
  title={A programmable hyper-dimensional processor architecture for human-centric IoT},
  author={Datta, Sohum and Antonio, Ryan AG and Ison, Aldrin RS and Rabaey, Jan M},
  journal={IEEE Journal on Emerging and Selected Topics in Circuits and Systems},
  volume={9},
  number={3},
  pages={439--452},
  year={2019},
  publisher={IEEE}
}

@article{adv-encoding-1-hv-design-eff-hdc,
  title={Hypervector design for efficient hyperdimensional computing on edge devices},
  author={Basaklar, Toygun and Tuncel, Yigit and Narayana, Shruti Yadav and Gumussoy, Suat and Ogras, Umit Y},
  journal={arXiv preprint arXiv:2103.06709},
  year={2021}
}

@article{adv-encoding-2-encoding-binarized-img,
  title={An encoding framework for binarized images using hyperdimensional computing},
  author={Smets, Laura and Van Leekwijck, Werner and Tsang, Ing Jyh and Latr{\'e}, Steven},
  journal={Frontiers in big data},
  volume={7},
  pages={1371518},
  year={2024},
  publisher={Frontiers Media SA}
}

@article{adv-encoding-3-hardware-aware-static-opt,
  title={Hardware-Aware Static Optimization of Hyperdimensional Computations},
  author={Yi, Pu and Achour, Sara},
  journal={Proceedings of the ACM on Programming Languages},
  volume={7},
  number={OOPSLA2},
  pages={1--30},
  year={2023},
  publisher={ACM New York, NY, USA}
}

@inproceedings{adv-encoding-4-tiny-hd,
  title={tiny-hd: Ultra-efficient hyperdimensional computing engine for iot applications},
  author={Khaleghi, Behnam and Xu, Hanyang and Morris, Justin and Rosing, Tajana {\v{S}}imuni{\'c}},
  booktitle={2021 Design, Automation \& Test in Europe Conference \& Exhibition (DATE)},
  pages={408--413},
  year={2021},
  organization={IEEE}
}

@inproceedings{static-encoding-8-distri-HD,
  title={DistriHD: a memory efficient distributed binary hyperdimensional computing architecture for image classification},
  author={Liang, Dehua and Shiomi, Jun and Miura, Noriyuki and Awano, Hiromitsu},
  booktitle={2022 27th Asia and South Pacific Design Automation Conference (ASP-DAC)},
  pages={43--49},
  year={2022},
  organization={IEEE}
}

@inproceedings{static-encoding-9-hdc-framework-image-descriptors-cvpr,
  title={Hyperdimensional computing as a framework for systematic aggregation of image descriptors},
  author={Neubert, Peer and Schubert, Stefan},
  booktitle={Proceedings of the IEEE/CVF conference on computer vision and pattern recognition},
  pages={16938--16947},
  year={2021}
}

@inproceedings{rosing-vision-hd,
author = {Asgarinejad, Fatemeh and Morris, Justin and Rosing, Tajana and Aksanli, Baris},
title = {VisionHD: Towards Efficient and Privacy-Preserved Hyperdimensional Computing for Image Data},
year = {2024},
isbn = {9798400706882},
publisher = {Association for Computing Machinery},
address = {New York, NY, USA},
url = {https://doi.org/10.1145/3665314.3670852},
doi = {10.1145/3665314.3670852},
booktitle = {Proceedings of the 29th ACM/IEEE International Symposium on Low Power Electronics and Design},
pages = {1–6},
numpages = {6},
location = {Newport Beach, CA, USA},
series = {ISLPED '24}
}

@inproceedings{trainable-hd-old-version,
  title={Efficient hyperdimensional learning with trainable, quantizable, and holistic data representation},
  author={Kim, Jiseung and Lee, Hyunsei and Imani, Mohsen and Kim, Yeseong},
  booktitle={2023 Design, Automation \& Test in Europe Conference \& Exhibition (DATE)},
  pages={1--6},
  year={2023},
  organization={IEEE}
}

@inproceedings{hetero-algorithm-hardware-co-design,
  title={Algorithm-hardware co-design for efficient brain-inspired hyperdimensional learning on edge},
  author={Ni, Yang and Kim, Yeseong and Rosing, Tajana and Imani, Mohsen},
  booktitle={2022 Design, Automation \& Test in Europe Conference \& Exhibition (DATE)},
  pages={292--297},
  year={2022},
  organization={IEEE}
}

@article{hetero-hpvm-hdc,
  title={HPVM-HDC: A Heterogeneous Programming System for Accelerating Hyperdimensional Computing},
  author={Arbore, Russel and Routh, Xavier and Noor, Abdul Rafae and Kothari, Akash and Yang, Haichao and Xu, Weihong and Pinge, Sumukh and Adve, Vikram and Rosing, Tajana and Zhou, Minxuan},
  journal={arXiv preprint arXiv:2410.15179},
  year={2024}
}

@inproceedings{gpu-hdtorch, series={ICCAD ’22},
   title={HDTorch: Accelerating Hyperdimensional Computing with GP-GPUs for Design Space Exploration},
   url={http://dx.doi.org/10.1145/3508352.3549475},
   DOI={10.1145/3508352.3549475},
   booktitle={Proceedings of the 41st IEEE/ACM International Conference on Computer-Aided Design},
   publisher={ACM},
   author={Simon, William Andrew and Pale, Una and Teijeiro, Tomas and Atienza, David},
   year={2022},
   month=oct, pages={1–8},
   collection={ICCAD ’22} }

@article{gpu-openhd,
  title={Openhd: A gpu-powered framework for hyperdimensional computing},
  author={Kang, Jaeyoung and Khaleghi, Behnam and Rosing, Tajana and Kim, Yeseong},
  journal={IEEE Transactions on Computers},
  volume={71},
  number={11},
  pages={2753--2765},
  year={2022},
  publisher={IEEE}
}

@inproceedings{fpga-imani-revisiting-hdc-fpga,
  title={Revisiting hyperdimensional learning for fpga and low-power architectures},
  author={Imani, Mohsen and Zou, Zhuowen and Bosch, Samuel and Rao, Sanjay Anantha and Salamat, Sahand and Kumar, Venkatesh and Kim, Yeseong and Rosing, Tajana},
  booktitle={2021 IEEE International Symposium on High-Performance Computer Architecture (HPCA)},
  pages={221--234},
  year={2021},
  organization={IEEE}
}

@inproceedings{pim-hdnn-rosing,
  title={Hdnn-pim: Efficient in memory design of hyperdimensional computing with feature extraction},
  author={Dutta, Arpan and Gupta, Saransh and Khaleghi, Behnam and Chandrasekaran, Rishikanth and Xu, Weihong and Rosing, Tajana},
  booktitle={Proceedings of the Great Lakes Symposium on VLSI 2022},
  pages={281--286},
  year={2022}
}

@INPROCEEDINGS{onlinehd,
  author={Hernández-Cano, Alejandro and Matsumoto, Namiko and Ping, Eric and Imani, Mohsen},
  booktitle={2021 Design, Automation \& Test in Europe Conference \& Exhibition (DATE)}, 
  title={OnlineHD: Robust, Efficient, and Single-Pass Online Learning Using Hyperdimensional System}, 
  year={2021},
  volume={},
  number={},
  pages={56-61},
  doi={10.23919/DATE51398.2021.9474107}}

\end{document}